\documentclass[review]{elsarticle}

\usepackage{lineno,hyperref}
\usepackage{algpseudocode}
\usepackage{algorithm}
\usepackage{caption}
\usepackage{amssymb}
\usepackage{makecell}
\usepackage{multirow}
\usepackage{mathtools}
\usepackage{amsmath}
\usepackage{tabularx}
\usepackage{subfigure}
\usepackage{caption}
\modulolinenumbers[5]

\providecommand{\algorithminput}[1]{
	~\\
	\textbf{Input:}\\
	\begin{tabularx}{\textwidth}{rl}#1\end{tabularx}
}
\providecommand{\algorithmoutput}[1]{
	~\\
	\textbf{Output:}\\
	\begin{tabularx}{\textwidth}{rl}#1\end{tabularx}
}

\newcommand{\Break}{\textbf{break}}

\journal{Journal of System and Software}

\begin{document}

\begin{frontmatter}

\title{GPIC- GPU Power Iteration Cluster}

\author[mymainaddress1,mymainaddress3]{Gustavo R.L Silva}
\author[mymainaddress2,mymainaddress3]{Rafael R. Medeiros}
\author[mymainaddress1]{Ant\^onio P. Braga}
\author[mymainaddress3]{Douglas A.G. Vieira}
\address[mymainaddress1]{
Electronic Engineering Department - UFMG,
Av. Ant\^onio Carlos, 6627,
Belo Horizonte, Minas Gerais, Brazil}
\address[mymainaddress2]{Computer Engineering Department - CEFET-MG,  Av. Amazonas, 7675, Nova Gameleira, Belo Horizonte, Minas Gerais, Brazil}
\address[mymainaddress3]{ENACOM, Professor Jos\'e Vieira de Mendon\c{c}a, 770, Belo Horizonte, Minas Gerais, Brazil}





\begin{abstract}
%
%
%
%
This work presents a new clustering algorithm, the GPIC, a Graphics Processing Unit (GPU) accelerated algorithm for Power Iteration Clustering (PIC). Our algorithm is based on the original PIC proposal, adapted to take advantage of the GPU architecture, maintining the algorith original properties. The proposed method was compared against the serial and parallel Spark implementation, achieving a considerable speed-up in the test problems. 

\end{abstract}


\end{frontmatter}

\linenumbers

\section{Introduction}
\nolinenumbers
The progressive incorporation of data collection and communication
abilities into consumer electronics is gradually
transforming our society, affecting labor relations, and creating
new social habits. Consumer electronics, as well as industrial
and commercial equipments are continuously incorporating
functions that go far beyond their basic purpose \cite{bi2014internet}.  


The synergies between these physical and computational
elements form the base of a profound transformation in
the global economy. This new scenario expands current data
acquisition and processing technology infrastructure, and significantly
enhances connectivity, communication, computation,
control technology and decision-making capabilities \cite{bi2014internet}. 


Many current real problems may
involve terabytes/petabytes of data with hundreds/thousands of
variables that tend to rise continuously if the current growth
rate is maintained \cite{hashem2015rise}. Exponential growth of data storage capacity
in the worldwide network of devices is forecasted for the
next years, therefore, new adaptive data-driven algorithms that
consider the computational aspects of big data problems are
demanded \cite{Chen2014Min}.

This may appear as a dilemma in the area, since it would be
expected that a higher ability of sampling would improve representativeness
and performance \cite{fan2014challenges}. However, the reality is that,
in spite of a higher representativeness, most current methods
are not computationally capable to deal with such an increasing
high dimension and volume \cite{gandomi2015beyond}. 

The new big data scenario requires the investigation of new methods for data summarization \cite{leskovec2014mining}, subspace representation \cite{parsons2004subspace}, clustering \cite{bateni2014distributed}, optimization, as well as pattern recognition and forecasting with lower computational complexity. The implementation of such methods requires scalable computing platforms capable to store and process massive and high dimensional data \cite{manyika2014big}, \cite{kaur2015novel}. 

Therefore, the need for real-time inference turns dedicated hardware solutions, such as GPUs (Graphics Processing Units), into an alternative to scale-up learning algorithms \cite{andrade2013g}, \cite{bekkerman2011scaling}. In general, the main challenges for data set clustering in this context are the following: 
first,	high dimensionality, due to the huge amount of sources generating and storing data; second,	large volume of data, since data collected by the sources can also be huge due to local storage capacity.

This paper extends the Power Iteration Cluster (PIC) algorithm \cite{lin2010power} to GPU, called here as GPIC. As presented in the results section, this implementation provides a considerable speed-up when compared with the serial and parallel PIC (Spark) implementation \cite{picSpark2016}.

\section{Cluster Analysis}

A cluster is a collection of data objects which are similar to each other within the same group, or dissimilar to the objects
in other groups \cite{jain1999data}.
The cluster analysis can be defined as: given a representation
of a set of data points, find \textit{k} groups based on a metric
groups such that the similarities of the objects of the same
group are high while the similarities of objects in different
groups are low \cite{jain1999data}.

It aims at revealing the intrinsic structure of a given data set by forming statistically significant groups. Therefore, the aim is to separate a set of objects into groups where elements of the same group are similar to each
other and the elements of different groups are has dissimilarities \cite{zaiane2002data}.

This analysis can be applied to: generate a compact
summary of data for classification, pattern discovery, hypothesis
generation and testing, outlier detection, dynamic trend
detection, among others .


The Cluster Analysis presents some requirements and challenges
\cite{jain2010data}:
\begin{itemize}
\item Quality: ability to deal with different types of attributes such as numerical, categorical, text, multimedia, networks, and mixture of multiple types, discovery of clusters with arbitrary shape and ability to deal with noisy data;
\item Scalability: clustering all the data instead of only on
samples, high dimensionality and incremental or stream
clustering and insensitivity to input order;
\item Constraint-based clustering: user-given preferences or constraints like domain knowledge, user queries, interpretability and usability.
\end{itemize}

\subsection{Graph-based methods}
A graph-based methods first constructs a graph or hypergraph and then apply a clustering algorithm to a given partition \cite{gan2007data}. A link-based clustering algorithm can also be considered as a graph-based one, since the links between data
points can be considered as links between the graph nodes \cite{gan2007data}.
Typical graph-based clustering algorithms are Chameleon \cite{karypis1999chameleon}, CACTUS \cite{ganti1999cactus}, ROCK \cite{guha1999rock}, Spectral Cluster \cite{ng2002spectral}.

\subsubsection{Spectral Cluster}

Given a dataset $S$ with $n$ data points defined in $R^m$, it is defined an undirected and weighted graph $G$,  $A$ is the affinity matrix of $G$, $D$ is the diagonal degree of the matrix $d_{ii}=\sum^{n}_{j}a_{ij}$ , and $W= D^{-1}A$ is a normalized affinity matrix.

To improve computational cost of Spectral Cluster algorithms, some authors have proposed new methods.  Xiao et al.\cite{cai2013multi}, proposed a new robust large-scale multi-view clustering method to integrate heterogeneous representations of large scale data.

Donghui Yan et al.\cite{yan2009fast}, proposed a general framework for fast approximate spectral clustering in which a distortion-minimizing local transformation is first applied to the data. This framework is based on a theoretical analysis that provides a statistical characterization of the effect of local distortion on the mis-clustering rate \cite{yan2009fast}.

Another approach is proposed by Frank and Willian \cite{lin2010power}  to find a very low-dimensional embedding of a dataset using truncated power iteration on a normalized pair-wise similarity data matrix. This embedding turns out to be an effective cluster indicator, consistently outperforming widely used spectral methods such as NCut on real datasets \cite{lin2010power}. More details about PIC algorithm will be described on Methodology section.
%
%
%
%

\section{Clustering data using GPUs}
Recent advances in consumer computer hardware makes
parallel computing capability widely available to most users.
Applications that make effective use of the so-called Graphics
Processing Units (GPU) have reported significant performance
gains \cite{owens2008gpu}.

In CUDA (Compute Unified Device Architecture ) model, GPU is regarded as a co-processor which
is capable of executing a great number of threads in parallel.
A single source program includes host codes running on CPU (Central Processing Unit)
and also kernel codes running on GPU. Compute-intensive and
data-parallel tasks have to be implemented as kernel codes such a way 
 to be executed on GPU \cite{li2010speeding}.

In \cite{Nickolls2008}, the authors present an interesting study on how
several data mining applications can be implemented using
GPU in CUDA architecture \cite{garland2008parallel}. 
In \cite{farivar2008parallel}, a GPU version of k-means
algorithm is presented. Another approach that uses k-means over GPUs is presented in \cite{wu2009clustering}, in this approach the authors applied k-means algorithm on a synthetic data set with a one billion data points and 2 dimensions. The GPU-version took about 26 minutes, and the serial version of the code took 6 days.

Jianqiang et al.\cite{dong2013accelerating}, presented a GPU accelerated BIRCH  (Balanced Iterative Reducing and Clustering using Hierarchies) version that can be up to 154 times faster than the CPU version with good scalability and high accuracy.

In \cite{andrade2013g} the authors present the G-DBSCAN, a GPU parallel version of the DBSCAN (Density-Based Spatial Clustering of Applications with Noise) clustering algorithm. This algorithm using GPUs, can be over 100 times faster than its sequential version using CPU.

\section{Methodology}

\subsection{Power Iteration Clustering: An Overview}
The PIC algorithm is a Spectral clustering technique. This algorithm proposes a method for computing the largest eigenvector of a matrix by the Power Iteration Method \cite{lanczos1950iteration}.

The Power Iteration Method \cite{lanczos1950iteration} can be described like that,  let $W$ be a diagonalizable $n \times n$ matrix with dominant eigenvalue $\lambda_{1}$. So, there is a nonzero vector $x_{0}$ such that the sequence of vectors $x_{m}$ \cite{lanczos1950iteration}:
\begin{equation}
    \label{simple_equation}
    x_{1} =Wx_{0},x_{2}=Wx_{1},x_{3}=Wx_{2},...,x_{k}=Wx_{t-1}
\end{equation}
tends to a dominant eigenvector of $W$.
\begin{equation}
    \label{simple_equation}
    |\lambda_{1}| > |\lambda_{2}| \geq \lambda_{3}| \geq ... \geq |\lambda_{n}|
\end{equation}
Let $v_{1}$, $v_{2}$, ... , $v_{n}$ the corresponding eigenvectors. As $v_{1}, v_{2}, .... v_{n}$ are linearly independent, they form a basis of $\mathbb{R}^{n}$. Consequently, we can write $x_{0}$ as a linear combination of these  eigenvectors:
\begin{equation}
    \label{simple_equation}
    x_{t}=W^{t}x_{0}, t \geq 1
\end{equation}
For
\begin{equation}
\begin{array}{lcl} W^{t}x_{0} = c_{1}\lambda_{1}^tv_{1} + c_{2}\lambda_{2}^tv_{2}+ ... + c_{n}\lambda_{n}^{t}v_{n} \\ 
= \lambda_{n}^{1}\left (c_{1}v_{1}+ c_{2}\left (\frac{\lambda_{2}}{\lambda_{1}}  \right )^t v_{2} + ... + c_{n}\left (\frac{\lambda_{n}}{\lambda_{1}} \right ) ^t v_{n} \right ) \end{array}
\end{equation}

The fact that $\lambda_{1}$ be the dominant eigenvalue means that each of the fractions, is less than 1 in absolute value. Like this,

\begin{equation}
    \label{simple_equation}
    \left (\frac{\lambda_{2}}{\lambda_{1}}  \right )^t, \left (\frac{\lambda_{3}}{\lambda_{1}}  \right )^t, \left (\frac{\lambda_{n}}{\lambda_{1}}  \right )^t,
\end{equation}

a sequence that tends to zero as m which t $\rightarrow \infty$

Let $\lambda_{1} \neq 0 $ e $v_{1} \neq 0$, $x_{k}$ approximates a multiple of $v_{1}$, that is an eigenvector corresponding to $\lambda_{1}$, if $c_1 \neq 0$.

In the PIC \cite{lin2010power} algorithm the Power Iteration Method \cite{lanczos1950iteration} performs the update step:

\begin{equation}
    \label{simple_equation}
    v_{t+1}=\frac{Wv_{t}}{\|Wv_{t}\|_{1}}
\end{equation}

A serial version of PIC \cite{lin2010power} is presented in \ref{alg:picSerialImplementation}.

\begin{algorithm}[H]
\caption{Power Iteration Cluster \cite{lin2010power}}
\label{alg:picSerialImplementation}

\algorithminput{
$W$ &  Row-normalized affinity matrix  \\ 
$k$ & Number of clusters \\
$v_0$ & Initial vector\\
$\epsilon$ & Precision}
\vspace{-0.3cm}
\algorithmoutput{$C$ & Clusters $C_{1},C_{2}, ... ,C_{k}$}
\vspace{-0.5cm}
\begin{algorithmic}[1]
\State $\delta_0 \gets v_0 $ \Comment{Initialize the variation of eigenvector}
\For{$t= 0, 1, 2, ...$}
\State{$v_{t+1} \leftarrow \frac{Wv_{t}}{\parallel Wv_{t} \parallel_{1}}$ \Comment{update the eigenvectors of $W$}
\State  $\delta_{t+1} \leftarrow |v_{t+1} - v_{t}|$}  \Comment{ Update the variation of the previous and current eigenvector}
\If{{$|\delta_{t+1}-\delta_{t}| \leq \epsilon$}} \Comment{Stop criteria}
\State \Break \Comment{Terminate the estimation of eigenvectors}
\EndIf
\EndFor
\State  $C \gets $ $k$-means of $v_{t}$  \Comment{Cluster eigenvalues}
\State \Return $C$
\end{algorithmic}
\end{algorithm}

\subsection{Our Approach GPIC - Gpu Power Iteration Clustering}
The authors of the PIC \cite{lin2010power} algorithm have shared an implementation in MATLAB language \cite{Frank2015}. With this code, it was conducted an experiment to evaluate the PIC behavior for a moderate amount of data. We used two synthetic datasets, two moons and three circles with $n$= 15,000; 30,000 and 45,000 data points.

The results of the experiment indicate that the main bottleneck in the PIC \cite{lin2010power} is computing pair-wise distance/similarity for all data points $O(n^{2})$ (in the worst case). On average consumes 88.61\% of the PIC total time. The results of the experiment are presented in Table \ref{tab:tableDetailTimes}.

%
%
%
%
\begin{table}[H]
\centering
\caption{Runtime (seconds) consume of PIC algorithm.($m$=2)}
\label{tab:tableDetailTimes}
\begin{tabular}{|c|c|c|c|c|c|}
\hline
Dataset   & n                    & A size                  & A Time {[}s{]} & PIC total time {[}s{]} & \% total time \\ \hline
2 moons   & \multirow{2}{*}{15k} & \multirow{2}{*}{3.85GB} & 2.534          & 2.548                  & 99.45\%       \\ \cline{1-1} \cline{4-6} 
3 circles &                      &                         & 2.194,27       & 2.363,64               & 92.83\%       \\ \hline
2 moons   & \multirow{2}{*}{30k} & \multirow{2}{*}{16GB}   & 18.666,39      & 21.680,05              & 86.09\%       \\ \cline{1-1} \cline{4-6} 
3 circles &                      &                         & 18.064,53      & 21.091,17              & 85.64\%       \\ \hline
2 moons   & \multirow{2}{*}{45k} & \multirow{2}{*}{36.5GB} & 97.966,50      & 103.778,10             & 94.39\%       \\ \cline{1-1} \cline{4-6} 
3 circles &                      &                         & 62.228,62      & 84.977,15              & 73.22\%       \\ \hline
\end{tabular}
\end{table}

To solve this bottleneck we propose a GPIC (GPU Power Iteration Clustering) approach. GPIC creates a set of CUDA kernels and can be called one after another to perform different steps of the PIC algorithm. In order to process large volumes of data, GPIC approach divided the data in chunks and these chunks are copied to the \textit{device} iteratively. 

\begin{figure}[H]
\begin{center}
\includegraphics[width=7cm]{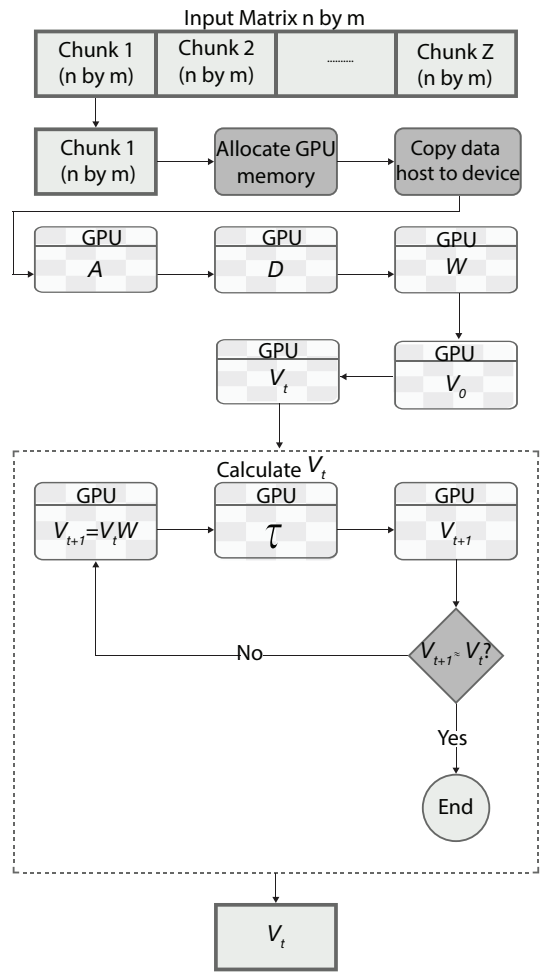}
\caption{GPIC execution flow.} 
\label{fig:gpicFlow}
\end{center}
\end{figure}

GPIC Algorithm is described in \ref{alg:gpic} and the figure \ref{fig:gpicFlow} presents the GPIC Algorithm execution flow. The main focus of GPIC is the Affinity Matrix and the Power Iteration step.

The first kernel \textit{\textbf{AffinityMatrix($S,p$)}} is launched to evaluated the Affinity Matrix $A$, where p is the number of threads.  The $A$ matrix is symmetric of size $n\times n$, as $n$ is the size of the input. When launched this kernel has $p$ threads and each thread is responsible for calculating a set of rows of the matrix $A$, $n_r$. Thus, each element of the matrix $A$ is indexed from $id$ of each thread and current index $i$. This step has $O(n^2/p)$ computational complexity, each CUDA core calculate parts of the $A$ matrix, where p is the number of threads. 

The second kernel \textit{\textbf{RowSum($A,p$)}}
is launched in order to sum the lines of $A$ matrix and the result of this operation is stored in the vector $D$. This step has $O(n/p)$ computational complexity.

The third kernel \textit{\textbf{NormMatrix($A, D, p$)}} is launched to normalize the affinity matrix $A$, and the results is stored in $W$ matrix. This step has $O(n^2/p)$ computational complexity.

\begin{figure}[H]
\begin{center}
\includegraphics[scale=0.44]{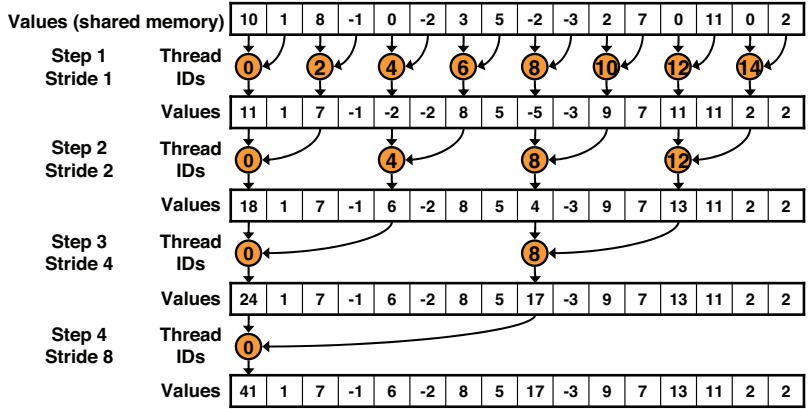}
\caption{Parallel Reduction: Interleaved Addressing \cite{NvidiaReduce2015}.} 
\label{fig:reductionKernel}
\end{center}
\end{figure}

The fourth kernel \textit{\textbf{Reduction($D,p$)}} is launched to calculate the sum of the vector $D$. In the step 9 of GPIC algorithm \ref{alg:gpic}, this operation is invoked again and the result is stored in $\tau$. This kernel has $O(n/p+log(n))$ computational complexity. Sequential reduction occurs in O(n) and in a thread block where $N=P$ (processors) we have $O(log N)$ complexity. We used sequential addressing pattern suggests by Nvidia  \cite{NvidiaReduce2015} because is conflict free. Figure \ref{fig:reductionKernel} describe how this reduction kernel works.

%
%
%
%

The fifth kernel \textbf{Norm($v_{t+1}, \tau,p$)} is launched to normalize the vector $D$, and results are stored in the $v_{t}$. In the GPIC algorithm step 10\ref{alg:gpic}, this operation is launched again to calculate the new values of $v_{t}$. This step has $O(n/p)$ computational complexity.
%
%
%
%

The last kernel \textit{\textbf{Multiply($v_{t}, W$)}} is launched to multiply the matrix $W$ by the vector $v_t$. This step has $O(n^2/p)$ computational complexity.

\begin{algorithm}[H]
\caption{GPU Power Iteration Cluster}
\label{alg:gpic}

\algorithminput{
$A$ &  Affinity Matrix  \\ 
$k$ &  Numbers of clusters \\
$v_0$ & Initial vector\\
$p$ & Number of threads\\
$\epsilon$ & Precision}
\vspace{-0.3cm}
\algorithmoutput{$C$ & Clusters $C_{1},C_{2}, ... ,C_{k}$}
\vspace{-0.5cm}
\begin{algorithmic}[1]

\State $A$ $\leftarrow$ \textbf{AffinityMatrix($S,p$)} \Comment{Affinity matrix $A$ Kernel}
\State $D$ $\leftarrow$ \textbf{RowSum($A,p$)} \Comment{Sum lines of  $A$ kernel}
\State $W$ $\leftarrow$ \textbf{NormMat($A,D,p$)} \Comment{Normalize $A$ kernel}
\State $v_0$ $\leftarrow$ \textbf{Reduction($D,p$)} \Comment{ Sum $D$ kernel}
\State $v_{t}$ $\leftarrow$ \textbf{Norm($D,v_0,p$)} \Comment{Normalize $D$ kernel}
\State $\delta_0 \gets v_0 $ \Comment{Initialize the variation of eigenvector}
\For{$t= 0, 1, 2, ...$}
\State $v_{t+1}$ $\leftarrow$ \textbf{Multiply($v_{t},W,p$)} \Comment{Multiply $W$ by $v_{t}$ Kernel}
\State $\tau$ $\leftarrow$ \textbf{Reduction($v_{t+1},p$)} \Comment {Sum $v_{t+1}$ Kernel}
\State $v_{t+1}$ $\leftarrow$ \textbf{Norm($v_{t+1},\tau,p$)} \Comment{Update the eigenvectors of $W$ Kernel}
\State $\delta_{t+1} \leftarrow |v_{t+1}-v_{t}|$ \Comment{ Update the variation of the previous and current eigenvector}
\If{{$|\delta_{t+1}-\delta_{t}| \leq \epsilon$}} \Comment{Stop criteria}
\State \Break \Comment{Terminate the estimation of eigenvectors}
\EndIf
\EndFor
\State  $C \gets $ $k$-means of $v_{t}$  \Comment{Cluster eigenvalues}
\State \Return $C$
\end{algorithmic}
\end{algorithm}

This GPU implemented PIC method converges to exact the same result of the original serial method, since the multi-thread explores the fact that the operations are independent.

\section{Results}
\subsection{Experiment I - Benchmark running time and speedup metrics}

In this section the results of the proposed GPIC algorithm is compared to serial version of the PIC \cite{lin2010power} and the parallel version of PIC implemented in Spark \cite{picSpark2016}. 

Experiments were conducted on a server containing two Intel
Xeon E5-2620 (2 GHz, totally 24 cores) with 64 GB RAM and one k40m model NVIDIA card. This card k40m has 12 GB GDDR5 SDRAM and 2880 CUDA cores with 745 MHz. 

Performance was evaluated using  two synthetic datasets (two moons and three circles) with the matrices ranging in size from 15.000 $\times$ 15.000 (3.85GB), 30.000$\times$30.000 (16GB) and 45.000$\times$45.000 (36GB),
and each test case was run ten times. The average results are presented.

Table \ref{tab:table2moons3Circles} shows the running time and speedup of PIC, PIC parallel (Spark implementation)\cite{picSpark2016} and GPIC, varying volume data size.

\begin{center}
\begin{table}[H]
\centering
\caption{Runtime (in seconds) and speedup comparison of PIC (serial version)\cite{lin2010power}, PIC parallel (Spark implementation) \cite{picSpark2016} and GPIC method on two synthetic datasets. The parameters for all experiments are maxiterations=3, precision=0.00001/n, m=2, and cosine similarity function.}
\label{tab:table2moons3Circles}
\begin{tabular}{|c|c|c|c|c|c|c|c|}
\hline
Dataset   & $n$                       &  \thead{$A$ \\size}& \thead{PIC\\serial\\ Time [S]} & \thead{Parallel \\  PIC\\ Time [S]} & \thead{GPIC \\ Time [S]} & \thead{Speedup \\ GPIC\\x\\PIC}   & \thead{Speedup \\ Parallel \\ PIC \\x \\PIC} \\ \hline
2 moons & \multirow{2}{*}{15k} &  \multirow{2}{*}{3.85G} & 2.548 & 467 & 3,99 & 638,60 & 5,44 \\ \cline{1-1} \cline{5-8}
3 circles &                         & & 2.363          & 320  & 4,03 & 586,51 & 7,38   \\ \cline{1-1} \cline{4-8}
2 moons & \multirow{2}{*}{30k} &  \multirow{2}{*}{16G} & 21.680  & 1.074  & 18,00 & 1.204,45 &  20,19  \\ \cline{1-1} \cline{5-8}
3 circles &                         & & 21.091          & 1.063   & 18,03 & 1.169,78 & 19,84 \\ \cline{1-1} \cline{4-8}
2 moons & \multirow{2}{*}{45k} &  \multirow{2}{*}{36.5G}  &103.778          &  23.723 & 45,07 & 2.302,60 &  4,36 \\ \cline{1-1} \cline{5-8}
3 circles &                         & & 84.977          & 22.976 & 45,90 & 1.851,35 &  3,69  \\ \hline
\end{tabular}
\end{table}
\end{center}

\begin{figure}[H]
\begin{center}
\subfigure[Gpic $\times$ PIC (Y axes in log scale)]{\includegraphics[scale=0.4]{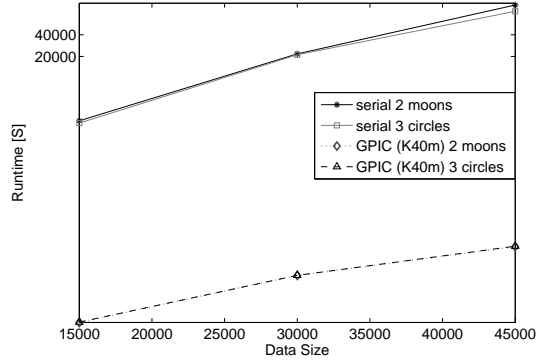}}
\subfigure[Parallel Pic on Spark $\times$ PIC (Y axes in log scale)]{\includegraphics[scale=0.4]{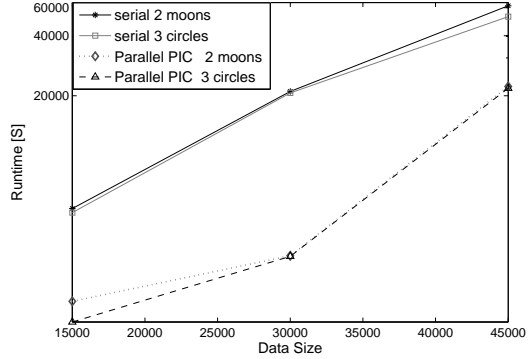}}
\caption{Pic $\times$ GPIC and Pic $\times$ Parallel Pic on Spark }. 
\label{fig:plotResults}
\end{center}
\end{figure}

This experiment results indicate a significant speedup gain for the GPIC approach which on average was 1,292.19 times faster than the original PIC in this two synthetic datasets. The parallel version of PIC implemented in Spark \cite{picSpark2016}  get a speedup of only 10.15 times on average when compared with original serial PIC version. The figure \ref{fig:plotResults} presents the run time with varying data sizes.

%
%
%
%

\subsection{Experiment II - Subsampling to reduce input data matrix}
One of the most important point observed in the previous experiment was that the increasing the amount of data can saturate both resources CPU and GPU memory.

To treat this problem, it was conducted another experiment to sample a portion of data and check the quality of the cluster with an external cluster validation indexes Adjusted Rand Index \cite{HubertLawrence1988} and Jaccard Index \cite{jaccard1908nouvelles}.

This experiment considers four synthetic datasets (Cassine, Gaussian, Shapes and Smiley) with $n$ equals 45.000 samples \ref{fig:4_datasets}.

\begin{figure}[H]
\begin{center}
\includegraphics[scale=0.7]{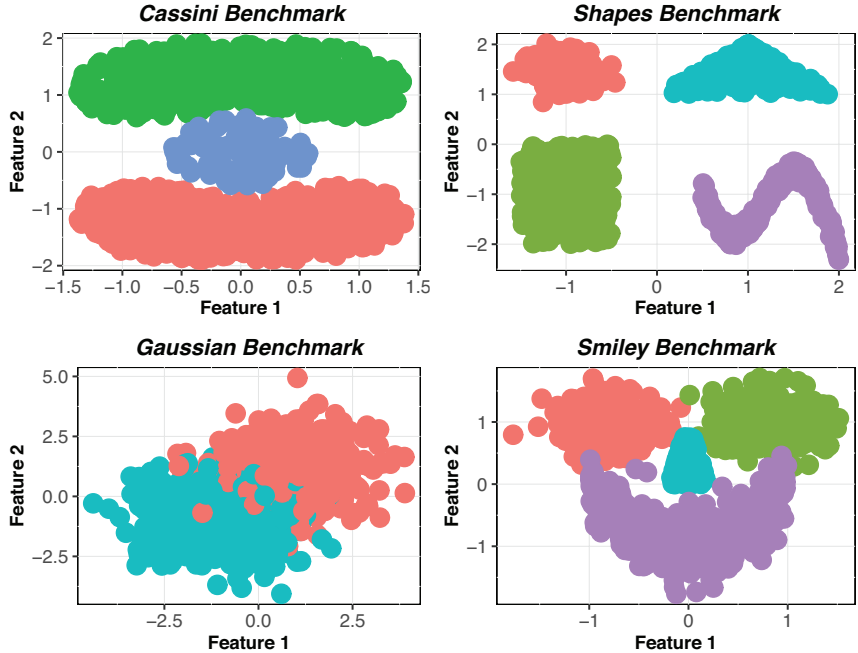}
\caption{Four synthetic datasets used on experiment II.} 
\label{fig:4_datasets}
\end{center}
\end{figure}

Our approach to subsample the data considered the strategies bellow:
\begin{itemize}
\item size of data varies from 0.01\% to 0.09\% and  0.1\%,to 0.9\% accounting eighteen data samples;
\item balanced classes;
\item run ten times and collect the mean and standard deviation of Adjusted Rand Index \cite{HubertLawrence1988} and Jaccard Index \cite{jaccard1908nouvelles}.
\end{itemize}

Through  figure \ref{fig:4_datasets} it can be concluded that, for this four synthetic datasets the strategy to reduce the amount of data used in the $X$ input matrix did not have a significant impact on the quality of GPIC when evaluated from the perspective of the Adjusted Rand Index \cite{HubertLawrence1988} and Jaccard Index \cite{jaccard1908nouvelles}. 

\begin{figure}[H]
\begin{center}
\includegraphics[scale=0.59]{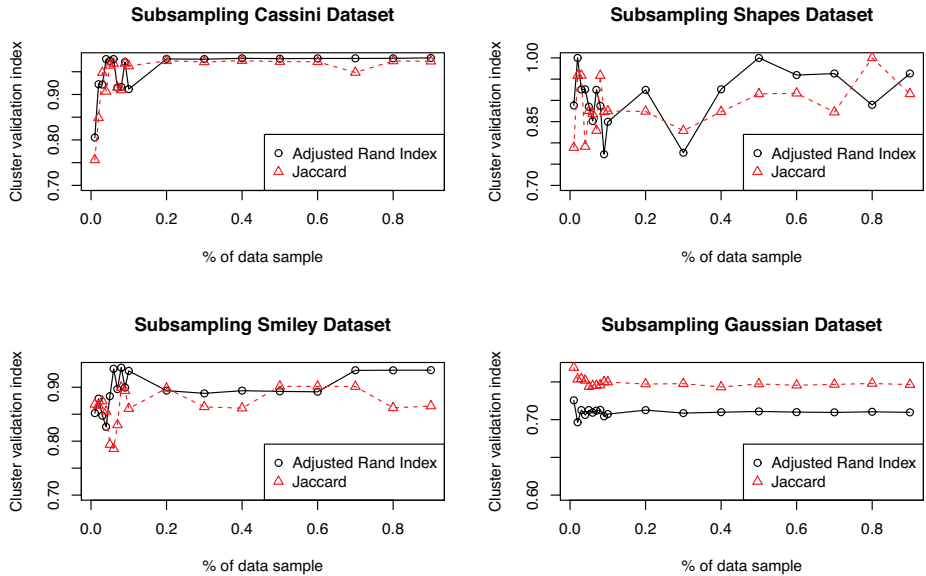}
\caption{Four synthetic datasets used on experiment II.} 
\label{fig:4_datasets}
\end{center}
\end{figure}

\section{Conclusion}

This paper presents a new clustering algorithm, the GPIC, a GPU accelerated algorithm for Power Iteration Clustering. Our algorithm is based on the original PIC proposal, adapted to take advantage of the GPU architecture. The proposed method was compared against the serial and parallel Spark implementation, achieving a considerable speed-up in the test problems. Experimental results demonstrated that GPIC implementation has an excellent scalability.

It was analyzed the impact on cluster quality when the number of samples was reduced. With this experiment it can be seen that the cluster quality had no significant variations with data reduction. This experiment was very important to indicated a way when the data does not fit on CPU and GPU memory.

For future work we plan to use a version of GPIC algorithm on multiple GPUs boards to process big data problems.  Multiple GPUs can compute faster (more GPUs equals faster time to a solution),  larger (more GPUs means more memory for larger problems) and cheaper (more GPUs per node translates to less overhead in money, power and space).

We think that on multi-GPU boards and sampling approaches to reduce the amount of data input, GPIC algorithm can achieves linear speedup, and significant performance improvement for a big datasets.

\section*{References}

\bibliography{elsarticle-template}

\end{document}